\documentclass{llncs}       
\usepackage{color}
\usepackage{hyperref}
\usepackage{lineno}
\usepackage{bsymb}

\definecolor{pblue}{rgb}{0.13,0.13,1}
\definecolor{pgreen}{rgb}{0,0.5976,0}
\definecolor{pred}{rgb}{0.6,0,0}
\definecolor{pgrey}{rgb}{0.75,0.75,0.75} 
\definecolor{darkgreen}{rgb}{0,0.8,0.3} 
\definecolor{darkgrey}{rgb}{0.5,0.5,0.5}

\def \inv       {^\sim}

\newcommand{\ebkeyw}[1]{\textcolor{pred}{\textsf{#1}}}
\newcommand{\ebtag}[1]{\textcolor{pgrey}{\textsf{\bf #1}}}
\newcommand{\ebcmt}[1]{\textcolor{pgrey}{\textsf{\it #1}}}
\newcommand{\ebcode}[1]{\textsf{#1}}
\newcommand{\ebcons}[1]{\textcolor{pgreen}{\textsf{#1}}}

\begin{document}

\title{Improving the Reliability of Mobility Applications}

\titlerunning{Improving the Reliability of Mobility Applications}    

\author{N{\'e}stor Cata{\~n}o}

\institute{ 
\email{nestor.catano@gmail.com} 
}

\date{}

\maketitle

\begin{abstract}
  The Android platform \cite{Android} was introduced by Google in 2008
  as an operating system for mobile devices. Android's SDK
  \cite{AndroidDev} provides a wide support for programming and
  extensive examples and documentation. Reliability is an increasing
  concern for Smart Phone applications since they often feature
  personal information and data. Therefore, techniques and tools for
  checking the correct behaviour of apps are required. This paper
  shows how the Event-B method can be used to reason and to verify the
  design of Android apps and how this can be used to document
  implementation decisions. Our approach consists in modelling the
  core functionality of the app in Event-B and using the evidence
  shown by the Proof Obligations generated to reason about the design
  and the implementation of the app. Although we don't propose a novel
  approach, we prove that heavyweight Formal Methods (FMs) techniques
  with Event-B can effectively be used to support the development of
  correct Android apps. We present a case study in which we design the
  core functionality of WhatsApp in Event-B, we encode it over three
  machine refinements modelling basic functionality (chatting,
  deleting content, forwarding content, deleting a chat session,
  etc.), read and unread status of chat sessions, and implementation
  details, respectively. We report and discuss on underlying
  challenges in the design and implementation of the core
  functionality.

\end{abstract}


\section{Introduction} 
\label{sec:intro}

Mobile phones have never been more popular and exciting for
programmers. Android operating system \cite{Android} was first
released by Google in October 2008, and it's today an ongoing
development effort led by the Open Handset Alliance (OHA) and
Google. On September 2013 Google declared that more than one billion
of Android devices were in use in the world, with over one million of
Android apps published in the Google Play store. Android is built on
top of an open source framework that features powerful libraries for
mobile development, primarily designed for touchscreen smartphones and
tablets. 
 One of the most appealing aspects of Android is
that it allows developers to use on a smartphone services like Gmail
or Calendar which are typically used online. Reliability of Smart
Phone apps is a significant concern as they often manipulate personal
information and data. This is exacerbated by the fact that the
increase of manipulated data by Smart Phones also brings opportunities
for privacy and security breaches. Therefore, sound techniques for the
development of correct Android apps are required. 

In spite of some misconception about their cost-effectiveness (results
do not outweigh the investment in time and money), Formal Methods
(FMs) have proven their potential to dramatically increase the quality
of software systems as conceived and developed by the IT industry as
shown in various case studies presented by the author
\cite{smart:scp:05,CatanoH:02,SCP:Pulse:14,ERP:EB:15}. This paper
discusses about the use of heavyweight FMs techniques to check the
design of Android applications. We use the Event-B formalism
\cite{Abrial:Ref:07,Abrial:EB:Book:2010} to model the core
functionality of the Android app. Event-B language is based on
set-theory and predicate logic and it's that what makes it of great
value as it can be used by theorem provers to reason about underlying
properties. We present a case study in which we verify the design of
WhatsApp formally in Event-B using the Rodin IDE
\cite{rodin:plat}. WhatsApp is a popular freeware instant messenger
service for Smart Phones, available from Google Play Store
(\url{https://www.whatsapp.com/android/}).  We adopt a Software
Engineering (SE) approach to reason about the design of WhatsApp,
starting by a discussion about its software requirements, their
formalisation in Event-B, the verification of underlying Proof
Obligations (POs) with Rodin, and the use of POs for design and
implementation decisions,



The contributions of this paper are two-fold. ($i.$) We demonstrate
how the use of heavyweight FMs techniques with Event-B can be
employed to verify the design of the core functionality of an Android
app, and how implementation decisions can be made based on the
evidence shown by the Event-B modelling of the Android app. ($ii.$)
FMs techniques and languages such as Event-B have traditionally been
used to develop critical systems. This paper is unique in showing how
discrete mathematics and program refinement techniques
with Event-B can effectively be used to support the development of
Android applications.

\section{Background}
\label{sec:back}

Event-B is based on \emph{Action Systems} \cite{as:back:91}, a
formalism describing the behaviour of a system by the (atomic) actions
that the system carries out. An Action System describes the state
space of a system and the possible actions that can be executed in it.
Event-B models are composed of \emph{contexts} and \emph{machines}.
Contexts define constants, uninterpreted sets and their properties
expressed as \ebkeyw{axioms}, while machines define variables and
their properties, and state transitions expressed as events. The
\ebkeyw{initialisation} event computes the initial state of a machine.
An event is composed of a \emph{guard} and an \emph{action}. The guard
(written between keywords \ebkeyw{where} and \ebkeyw{then}) represents
conditions that must hold in a state for the event to trigger. The
action (written between keywords \ebkeyw{then} and \ebkeyw{end})
computes new values for state variables, thus performing an observable
state transition. If the system reaches a state where no event guard
holds, it halts and is said to have \emph{deadlocked}. There is no
requirement that the system should halt, and indeed, most Event-B
models represent systems that run forever. If halting is desired, the
system can be modelled using \ebkeyw{convergent} events that
monotonically decrease the value of a natural number expression called
the machine \ebkeyw{variant}. Such events can only be triggered in
states where the value of the \ebkeyw{variant} is non-negative.
Additionally, the system may reach a state where the guards of more
than one event hold. In this situation, the system is said to be
\emph{non-deterministic}: Event-B semantics allows any of the events
whose guards are satisfied to be triggered.

In Event-B, systems are typically modelled via a sequence of
refinements. First, an abstract machine is written to verify whatever
correctness and safety properties are desired.  Refinement machines
are used to add more detail to the abstract machine until the model is
sufficiently concrete for hand or automated translation to
code. Refinement Proof Obligations are discharged (proven) to ensure
that each refinement is a faithful model of the previous machine, so
that all machines satisfy the correctness properties of the original.

\begin{figure}
{\internallinenumbers{\raggedright
\ebkeyw{machine} \ebcode{machine0} \ebkeyw{sees} \ebcode{ctx0}\\
~\\
\ebkeyw{variables}~\ebcode{user content chat active chatcontent}\\
~\\
\ebkeyw{invariants}\\
  ~\ebtag{@inv1} \ebcode{user} $\subseteq$ \ebcons{USER} \\
  ~\ebtag{@inv2} \ebcode{content} $\subseteq$ \ebcons{CONTENT} \\
  ~\ebtag{@inv3} \ebcode{chat} $\in$ \ebcode{user} $\rel$ \ebcode{user}~~\ebcmt{// chat sessions}\\
  ~\ebtag{@inv4} \ebcode{active} $\in$ \ebcode{user} $\pfun$ \ebcode{user}~~\ebcmt{// active chat session}\\
~\\
\ebkeyw{events}\\
  ~\ebkeyw{event}~\ebkeyw{initialisation}\\
   ~\ebkeyw{then}\\
      ~~\ebtag{@init1}~\ebcode{user :=  $\emptyset$}\\
      ~~\ebtag{@init2}~\ebcode{content := $\emptyset$}\\
      ~~\ebtag{@init3}~\ebcode{chat := $\emptyset$}\\
      ~~\ebtag{@init4}~\ebcode{active := $\emptyset$}\\
  ~\ebkeyw{end}\\
~\\
  ~\ebkeyw{event}~\ebcode{create-chat-session}\\ 
    ~\ebkeyw{any}~\ebcode{u1 u2}\\
    ~\ebkeyw{where}\\
      ~~\ebtag{@grd1}~\ebcode{u1 $\in$ user $\wedge$ u2 $\in$ user}\\
      ~~\ebtag{@grd2}~\ebcode{u1 $\mapsto$ u2 $\not\in$ chat}\\
    ~\ebkeyw{then}\\
      ~~\ebtag{@act1} \ebcode{chat := chat $\cup$ \{u1 $\mapsto$ u2\}}\\
      ~~\ebtag{@act2} \ebcode{active(u1) := u2}\\
  ~\ebkeyw{end}\\ 
  \ebkeyw{end}\\ 
}}
  \caption{A simplified WhatsApp abstract machine in the Event-B language.}
  \label{fig:evbmachine}
\end{figure}

Figure \ref{fig:evbmachine} presents a simplified version of an
Event-B model of WhatsApp. The \ebkeyw{initialisation} event starting
on line 12 gives initial values to the state (machine) variables. One
further event is shown: one that is triggered when any user creates a
chat session between two users \ebcode{u1} and \ebcode{u2} (Line 21
and 23). Guard \ebtag{@grd2} checks that the chat does not already
already. Action \ebtag{@act1} modifies the chat to contain the pair of
elements \ebcode{u1 $\mapsto$ u2}. Action \ebtag{@act1} makes the new
added chat active for \ebcode{u1}. The construct:
\[
\ebkeyw{any}~x~\ebkeyw{where}~G(s,c,v,x)~\ebkeyw{then}~v~:=~A(s,c,v,x)~\ebkeyw{end}
\]
specifies a \emph{non-deterministic} event that can be triggered in a
state where the guard $G(s,c,v,x)$ holds for some bounded value $x$,
sets $s$, constants $c$, and machine variables $v$. When the
event is triggered, a value for $x$ satisfying $G(s,c,v,x)$ is non-deterministically chosen
and the event action $v := A(s,c,v,x)$ is executed with $x$ bound to that
value. The correctness condition of the event requires that, for any
$x$ chosen, the new values of the state variables computed by the
action of the event maintain the invariant properties of the machine.

\subsection{Event-B Mathematical Notation}

Event-B provides a full battery of set and relation notation. Figure
\ref{fig:notation} shows some of the Event-B mathematical notation
used by Event-B. We use square brackets to apply (evaluate) a relation
to (over) a set of elements as mentioned above. For instance,
\ebcode{r[s]} applies relation \ebcode{r} to all the elements in set
\ebcode{s}. The result of \ebcode{r[s]} is a set of elements in the
range of relation \ebcode{r}. Event-B provides standard notations for
set union, intersection, difference, etc. Symbol $\times$ denotes the
cross product between two sets. The operator \ebkeyw{dom} returns the
domain of a relation, and \ebkeyw{ran} its range relation. The
expression \ebkeyw{id}\ebcode{[s]} denotes the identity relation over
a set of elements \ebcode{s}. Applying the forward composition
relation \ebcode{q;r} to an element \ebcode{a} in the domain of
relation \ebcode{q} returns a set of elements calculated as the result
of applying \ebcode{r} to \ebcode{q[\{a\}]}. When a relation
\ebcode{q} is a function, \ebcode{q[\{a\}]} should be used as
\ebcode{q(a)}. The domain restriction relation expression
\ebcode{s$\domres$r} restricts the domain of a relation \ebcode{r} to
(consider only elements in) a subset \ebcode{s} of its domain. The
range restriction relation expression \ebcode{r$\ranres$s} restricts
the range of relation r to consider only elements in a subset
\ebcode{s} of its range. Domain (range) subtraction is defined
similarly to domain (range) restriction, except that the elements in
the set s are disregarded rather than considered.

\begin{figure}[t]
\begin{tabular}{|l|l|l|l|} \hline
{\bf Syntax} & {\bf Name} & {\bf Definition} & {\bf Short Form} \\\hline
$q;r$                & forward  & $\{(x,z)~|~\exists{}y\cdot{} (x,y)\in{}q~\wedge$ & $q;r$\\
                        & composition            & $\quad\quad\quad\quad\quad\quad (y,z)\in{}r\}$ & \\ \hline
$\ebkeyw{id}[s]$  & identity relation   & $\{(x,y)~|~(x,y)\in{}s\times{}s~\wedge$        & $\ebkeyw{id}(s)$ \\ 
                         &                                & $\quad\quad\quad\quad\quad{}~x = y\}$ & \\ \hline
$s\domres{}r$  & domain restriction   & $\{ (x,y)~|~(x,y) \in r \wedge x\in s\}$            & $\ebkeyw{id}(s);r$ \\ \hline
$s\domsub{}r$ & domain subtraction & $\{ (x,y)~|~(x,y) \in r \wedge x\not\in s\}$     & $(\ebkeyw{dom}(r)\setminus{}s) \domres r$ \\ \hline
$r\ranres{}s$    & range restriction      & $\{ (x,y)~|~(x,y) \in r \wedge y\in s\}$            & $\ebkeyw{id}(s);r$ \\ \hline
$r\ransub{}s$   & range subtraction    & $\{ (x,y)~|~(x,y) \in r \wedge y\not\in s\}$     & $ r \ranres (\ebkeyw{ran}(r)\setminus{}s)$ \\ \hline
$r[s]$               & relational image       &  $\{ y~|~(x,y) \in r \wedge x\in s\}$     & $\ebkeyw{ran}(s \domres{}r)$ \\ \hline
$r\oplus{}q$     & relational overriding &  $\{ (x,y)~|~(x,y) \in
q~\vee$ & $q\cup(\ebkeyw{dom}(q)\domsub{}r)$\\
                       &                               &  $\;\;((x,y)\in{}r\wedge\neg\exists{}z\cdot(x,z)\in{}q)\}$ & \\\hline 
$r\inv$            & inverse relation         &  $\{ (x,y)~|~(y,x) \in  r \}$  & $r\inv$      \\ \hline
\end{tabular}
\caption{Basic Event-B mathematical notation.}
\label{fig:notation}
\end{figure}

\subsection{Event-B Relations and Functions}

Event-B relations are encoded as a set of pairs. A relation \ebcode{f}
with domain \ebcode{A} and range \ebcode{B} is denoted \ebcode{f: A
  $\rel$ B}. If \ebcode{f} is a function defined for all values of
\ebcode{A}, we say that \ebcode{f} is a total function, and we write
\ebcode{f: A $\tfun$ B}. If \ebcode{f} is defined for some values of
\ebcode{A}, we say that \ebcode{f} is a partial function, and we write
\ebcode{f: A $\pfun$ B}. If \ebcode{f} is a function such as no
element in the range of \ebcode{f} is associated with more than one
element in the domain of \ebcode{f}, then we say that \ebcode{f} is a
one-to-one or injective function, and we write \ebcode{f: A $\tinj$
  B}. If \ebcode{f} is a function whose range is \ebcode{B}, we say
that f is an onto or surjective function, and we write \ebcode{f: A
  $\tsur$ B}. If \ebcode{f} is both one-to-one and onto, we say that
\ebcode{f} is a bijection, and we write \ebcode{f: A $\tbij$ B}.

 \subsection{Rodin and EventB2Java}
 Rodin \cite{rodin:plat} is an Eclipse based platform that provides
 support to Event-B, for instance, for writing Event-B models,
 defining safety invariant properties, and for discharging POs using
 back-end provers. EventB2Java \cite{b2jml:tool:16} is a plugin of the
 Rodin IDE. It generates Java implementations for Event-B
 programs. EventB2Java translates a \emph{machine} as a Java class. In
 translating a machine, EB2Java not only considers the information
 provided by the machine, but also the \emph{contexts} the machine
 sees. Refinement machines are translated in the same way as abstract
 machines since Rodin properly adds abstract machine components to the
 internal representation of the refining machine. Refining and
 extending events (defined using \ebkeyw{refines} and
 \ebkeyw{extends}, respectively) are translated in the same manner as
 abstract events. Each \emph{event} is translated to a separate Java
 class. The translation of each event includes an object reference to
 the machine class. The translation of a standard event includes a
 \ebcode{guard\_evt} method that tests if the guard of the event
 \ebcode{evt} holds, and a \ebcode{run\_evt} method that models the
 execution of \ebcode{evt}.

\section{Software Development with Event-B}
\label{sec:met:eb}

Software development with Event-B starts with the definition of an
initial \emph{blueprint} of the system one wants to model. This
blueprint represents the future system implementation. Blueprints give
insight on some but not all the aspects of the future system. A
blueprint then goes through a series of stages called refinements
\cite{EventB07}. A blueprint refinement adds details to the
blueprint. Refinements provide a hierarchical organization of the
blueprints. The design of the initial system blueprint and its
subsequent refinements is based on the description contained on an
existing software requirements document. Each stage of the
organization of a blueprint serves a different purpose. At higher
levels, blueprints are used to state key system properties. At lower
levels, blueprints implement the system behaviour. It is crucial that
the initial blueprint and its refinements are consistent with each
other, and that they are coherent with respect to the system
specification. A refinement step generates Proof Obligations (POs)
expressed in predicate logic to assert that the blueprint refinement
is a refinement of the blueprint. That is, POs guarantee that the
blueprint and its refinements are models of the same system.

Event-B caters for two types of blueprint refinements,
\emph{horizontal refinement} (discussed above) and \emph{vertical
  refinement} \cite{abrial:we:can:09}. Horizontal refinement is also
called \emph{superposition} in literature. Horizontal refinements add state
transitions to the system or enrich existing transitions. The
horizontal refinement stage is complete when all the software
requirements are considered in the model. Through horizontal
refinement a blueprint (a machine) can:

\begin{itemize}
\item strengthen an event guard, 
\item add new event guards, 
\item add more actions to some events, or 
\item add more events. 
\end{itemize}

Vertical refinement is \emph{data refinement}. It does not add more
details to the system, but it transforms the model into something that
can easily be implemented. For instance, vertical refinement can
transform finite sets into Boolean arrays. A key aspect of a vertical
refinement is the definition of a \emph{gluing invariant} that bridges
the abstract state of the system to the concrete state of the system
by stating properties of the combined behaviour of both state
models. Although horizontal and vertical refinements can be combined
together in a single refinement step, a final vertical refinement
single step is typically realized with the aid of a code generation
tool such as EventB2Java
\cite{sttt:codegen:17,b2jml:tool:16,B2Jml:12}.








The definition of the most abstract machine above and all its
refinements are based on an existing software requirements
document. The Rodin tool provides support for Event-B and Event-B
model refinement definition \cite{EventB07}. Rodin generates safety
and consistency POs in each refinement stage. Rodin includes several
semi-automatic theorem provers that provide users assistance with
proof discharging.

\section{WhatsApp Software Requirements}
\label{sec:wp}

Software development with Event-B relies on the \emph{parachute}
strategy for software development. Development starts with an initial
abstract blueprint of the system in Event-B, and then, as the
\emph{paratrooper} descends, more details become clearer to him, and
so he's able to add them to the system description. There is
no automated mechanism or magical recipe that tells us how English
written functional or non-functional requirements must be ported to
Event-B. Nonetheless, one can stick to strict guidelines for writing
software requirements as described below.

Software requirements might be related to the static part of Event-B,
its dynamic part, or both. One needs to model the context first, which
is related to the static part of the machine. A machine context
typically includes constants, sets, and axioms that are used in the
abstract machine and its refinements to declare invariants that
typeset the machine variables. Machine variables are the static part
of a machine.

I elicited the software requirements by directly installing the
version of WhatsApp that is available from Android's store.
Requirements do not cater for the Web version of WhatsApp but only for
its Smart Phone version. They cater for the most basic functionality
of WhatsApp. When one writes software requirements in Event-B one
should write an abstract machine (model) first and then successively
write refinement machines \cite{EventB07}. For each refinement machine
Proof Obligations (POs) are to be discharged in the Rodin platform
\cite{rodin:plat} to ensure that each machine is a proper refinement
of the most abstract machines. Only once all the machines are written
and all the POs are discharged one can consider the underlying system
has completely been modelled.

Table \ref{tab:hierarchy} presents WhatsApp machines hierarchy, as
they will be modelled in Event-B. The abstract machine observes basic
functionality for chat sessions including the functionality for
creating a chat session, selecting or un-selecting a chat, chatting,
deleting content (text, video, photos), removing content, deleting a
chat session, muting and un-muting a chat, and broadcasting and
forwarding network content. The first machine refinement includes
functionality to check whether chat content has been read or not. The
second machine refinement adds implementation details, for instance,
it represents content as a sequence (rather than a set) of content
items. This is important for us because the graphical interface of a
chat session is implemented as an ordered sequence of content items
that reads from the beginning to the end. Additionally, it would help
us state a property that says that for any chat session, the chat
content as seen by one of the two chat members reads exactly the same
as it is seen by the other chat member. This is a \emph{safety
  invariant property}, it says that ``some desirable situation always
holds”, or, that “nothing bad happens''. This safety property in
particular is not a property of the Smart Phone version of WhatsApp.

\begin{table}
\begin{tabular}{|l|l|}\hline
\textbf{Machine}~~ & ~~\textbf{Observations} \\ \hline
~machine 0~~ & ~~Basic functionality for chat sessions~ \\ \hline
~machine 1~~ & ~~Read and unread status \\ \hline
~machine 2~~ & ~~WhatsApp's implementation \\ \hline 
\end{tabular}
\caption{WhatsApp's Event-B machines hierarchy}     
\label{tab:hierarchy}
\end{table}

I present WhatsApp's requirements as User Stories (US). Each software
requirement is checked against an acceptance criterion that has 3 main
components, a \ebkeyw{Given} part that describes when the
functionality may be triggered/executed (which depends on the internal
state of the system or program), a \ebkeyw{When} part that tells us
when the functionality is to be executed (which depends on the user's
decision), and a \ebkeyw{Then} part that tells how the state of the
system is changed when the functionality changes. US are typical of
Agile methodologies, yet I use them here since their structure fit
the structure of events in which the \ebkeyw{Given} part is encoded
through event guards, the \ebkeyw{When} part is the event itself that
is triggered, and the \ebkeyw{Then} part is encoded via event
actions. In writing the US, in general, I try to keep myself away
from the interaction user-interface and focus on the core
functionality that WhatsApp needs to provide. However, a simple
user-interaction may involve the working and interplaying of multiple
core functionality.

In what follows, Sections \ref{sub:basic} and \ref{sub:basic:inv}
present the basic functionality for chat sessions (first row in Table
\ref{tab:hierarchy}). Section \ref{machine0} presents the modelling of
that basic functionality in Event-B, and Section \ref{decisions0}
discusses design and implementation decisions related to the basic
functionality of WhatsApp. Second row in Table \ref{tab:hierarchy} is
not discussed in this paper. Section \ref{sub:impdet} presents
WhatsApp's functionality for the third row in Table \ref{tab:hierarchy},
Section \ref{machine2} discusses its modelling in Event-B, and
Section~\ref{decisions2} discusses related design and implementation
issues.

\subsection{Basic Functionality for Chat Sessions}
\label{sub:basic}

US-01 describes the functionality for creating a chat session between
Me and Another-User. The chat may not exist already. \\

\begin{center}
\begin{tabular}{|p{1.8cm}|p{9cm}|} \hline
US-01 & create-chat-session \\ \hline
Description & As a user, I want to create a chat session so that I can
communicate with Another-User \\ \hline
Acceptance &\ebkeyw{Given}: A chat session between Me and Another-User does not exist \\
 Criterion &\ebkeyw{When}: I decide to create chat session with Another-User\\
  &\ebkeyw{Then}: Chat session between Me and Another-User is created\\ \hline
\end{tabular}
\end{center}
~\\

US-02 describes the functionality for selecting a chat session. The
effect of having two Given conditions is the condition obtained as the
conjunction of both. \\

\begin{center}
\begin{tabular}{|p{1.8cm}|p{9cm}|} \hline
US-02 & select-chat \\ \hline
Description & As a user, I want to select a chat session so that I can start chatting with Another-User  \\ \hline
Acceptance &\ebkeyw{Given}: A chat session between Me and Another-User exists \\
 Criterion &\ebkeyw{Given}: A chat session between Me and Another-User is not active\\
  &\ebkeyw{When}: I select a chat session with Another-User\\
  &\ebkeyw{Then}: The chat session between Me and Another-User is made active\\ \hline
\end{tabular}
\end{center}
~\\

US-03 introduces the functionality used for Me to chat with
Another-User. Sent content is made available for both users Me and
Another-User. \\

\begin{center}
\begin{tabular}{|p{1.8cm}|p{9cm}|} \hline
US-03 & chatting \\ \hline
Description & As a user, I want to send some content during a chat
session with Another-User so that I can transmit some information
\\ \hline
Acceptance &\ebkeyw{Given}: Chat session with Another-User is active \\
Criterion  &\ebkeyw{When}: Content is produced and sent by Me \\
  &\ebkeyw{Then}: Content is made available to Me as well as to Another-User  \\ \hline
\end{tabular}
\end{center}
~\\

WhatsApp implements two different behaviours for
erasing exchanged content: ``Remove For Me'' and ``Remove For
Everyone''. If the sender of the content wants to remove some content,
he is offered the option to remove it from his chat or to remove it
from his chat and from the chat as seen by the user he's chatting
with. On the other hand, if the receiver of the content wants to
delete it, he can only do it from his chat. These two behaviours are
described by US-04a and US-04b, respectively. Erasing is always the
type of subtle functionality difficult to encode in logic as one can
easily break the machine invariants, for instance, if one erases
content from one side of the chat and not from the other, one would
break any invariant on the equivalence of content read by both users
of a chat session. One would then need to add an event guard (a
\ebkeyw{Given} condition) that prevents such behaviour or rephrase
the invariant properly. \\

\begin{center}
\begin{tabular}{|p{1.8cm}|p{9cm}|} \hline
US-04a & delete-content \\ \hline
Description & As a user, I want to delete some content exchanged with another user during a chat session so that I unclutter my chat  \\ \hline
Acceptance &\ebkeyw{Given}: Content exists  \\
Criterion  &\ebkeyw{When}: Me decides to delete the content he has received \\
  &\ebkeyw{Then}: Me's content is deleted \\ \hline
\end{tabular}
\end{center}
~\\

\begin{center}
\begin{tabular}{|p{1.8cm}|p{9cm}|} \hline
US-04b & remove-content \\ \hline
Description & As a user, I want to remove some content exchanged with another user during a chat session so that I unclutter my chat  \\ \hline
Acceptance &\ebkeyw{Given}: The content exists  \\
Criterion  &\ebkeyw{When}: Me decides to remove the content he has sent \\
  &\ebkeyw{Then}: The content is deleted from Me and anyone to whom Me has sent the content \\ \hline
\end{tabular}
\end{center}
~\\

Chat sessions and associated content can be deleted as well. What
would it happen with the content seen by Another-User if the session
between Me and Another-User is deleted. Will that content be deleted
from Another-User as well?  Deleting a chat session between Me and
Another-User does not delete the content as seen by Another-User,
regardless of who sent the content to whom, however, a remove-content
US exists that deletes the content both ways.\\

\begin{center}
\begin{tabular}{|p{1.8cm}|p{9cm}|} \hline
US-05 & delete-chat-session \\ \hline
Description & As a user, I want to delete a chat session with Another-User \\ \hline
Acceptance &\ebkeyw{Given}: A chat session between Me and Another-User exists \\
Criterion  &\ebkeyw{When}: I select to delete the only active chat session \\
  &\ebkeyw{Then}: The chat session is deleted as well as its associated content \\ \hline
\end{tabular}
\end{center}
~\\

When a chat session has been muted, communication between the two chat
users is disabled both ways. Nevertheless, communication can be
enabled later on.

\begin{center}
\begin{tabular}{|p{1.8cm}|p{9cm}|} \hline
US-06 & mute-chat \\ \hline
Description & As a user, I want to mute a chat session so that I can
prevent communication with and from Another-User \\ \hline
Acceptance &\ebkeyw{Given}: Chat session between Me and Another-User exists \\
Criterion  &\ebkeyw{When}: I select to mute a chat session \\
  &\ebkeyw{Then}: Chat session is muted and no communication from Me to the muted user or vice-versa is permitted \\ \hline
\end{tabular}
\end{center}
~\\

\begin{table}[t]
\begin{tabular}{|c|p{9.8cm}|} \hline
Number & Invariant \\ \hline
1 & Users are uniquely identified throughout the system. \\ \hline 
2 & Content is uniquely identified throughout the whole system. \\ \hline 
3 & Chat sessions are uniquely identified throughout the system.  \\ \hline 
4 & A chat session relates exactly two users. \\ \hline 
5 & Only one chat session maximum can be established between two users. \\ \hline 
6 & A chat session between two users may have a set of associated content available to either or both of them.  \\ \hline 
7 & Content is associated to a chat session only if one the users of the session has sent the content to the other user or vice-versa. \\ \hline 
8 & Active and muted are disjoint concepts. That is, it is never the case that the same system reaches a state in which user A muted user B and either is actively chatting with the other one. \\ \hline 
9 & Chat sessions are not symmetric. That is, the fact that user A has created a chat session so as to chat with user B, does not necessarily mean that user B has a created session so as to chat with user A. \\ \hline 
10 & Active chat sessions are no symmetric. That is, the fact that user A is actively chatting with user B does not necessarily mean that user B is actively chatting with user A. \\ \hline 
11 & It is never the case that chat content exists associated to a pair of users for which no chat session exists. \\ \hline 
12 & Several chat sessions can be created, but only one (or none) created chat session may be active per user. \\ \hline 
13 & Chat communication with a muted user is no feasible: no content exchange is feasible from or to a muted chat. \\ \hline 
\end{tabular}
\caption{Local invariants for \ebcode{machine0}}     
\label{tab:inv:machine0}
\end{table}

US-07 is about to re-establish communication between two users of a
muted chat. Only the user who muted the chat can unmute it.\\

\begin{center}
\begin{tabular}{|p{1.8cm}|p{9cm}|} \hline
US-07 & create-chat-session \\ \hline
Description & unmute-chat \\ \hline
Acceptance &\ebkeyw{Given}: Chat session between Me and Another-User is muted \\
 Criterion &\ebkeyw{Given}: I had muted the chat session previously \\
  &\ebkeyw{When}: I select to unmute a chat session  \\
  &\ebkeyw{Then}: Communication between Me and Another-User is re-established \\ \hline
\end{tabular}
\end{center}
~\\

US-08 and US-09 describe the situation whereby some content is sent to
a group of users; forwarding a content requires that respective chat
sessions between Me and the group of users exist, broadcasting creates
new chat sessions if they do not exist already.\\

\begin{center}
\begin{tabular}{|p{1.8cm}|p{9cm}|} \hline
US-08 & broadcast \\ \hline
Description & As a user, I want to broadcast a content to a group of users so that I can communicate with all of them quickly \\ \hline
Acceptance &\ebkeyw{Given}: Me wants to broadcast some content \\
Criterion  &\ebkeyw{When}: Me decides to broadcast the said content to Other-Users \\
  &\ebkeyw{Then}: The content is sent to Other-Users \\ \hline
\end{tabular}
\end{center}
~\\

\begin{center}
\begin{tabular}{|p{1.8cm}|p{9cm}|} \hline
US-09 & forward \\ \hline
Description & As a user, I want to forward a content to a group of users so that I can communicate with all of them quickly \\ \hline
Acceptance &\ebkeyw{Given}: Me wants to forward some content \\
Criterion  &\ebkeyw{Given}: Respective chats between Me and Other-Users exist\\
  &\ebkeyw{When}: Me decides to forward the said content to Other-Users \\
  &\ebkeyw{Then}: The content is sent to Other-Users \\ \hline
\end{tabular}
\end{center}
~\\

US-10 is the counterpart of US-02, unselecting a chat requires the chat to be active. \\

\begin{center}
\begin{tabular}{|p{1.8cm}|p{9cm}|} \hline
US-10 & unselect-chat \\ \hline
Description & As a user, I want to unselect a chat session so that I can chat with Another-User  \\ \hline
Acceptance &\ebkeyw{Given}: A chat session between Me and Another-User exists\\
Criterion  &\ebkeyw{Given}: A chat session between Me and Another-User is active \\
  &\ebkeyw{When}: Me wants to make session Another-User inactive \\
  &\ebkeyw{Then}: Chat session between Me and Another-User becomes inactive \\ \hline
\end{tabular}
\end{center}
~\\

Notice that \ebcode{select-chat} and \ebcode{unselect-chat} could have
been written without requiring the chat to be inactive or active,
respectively. Thinking about their final encoding, the two events can
eventually be encoded by adding a respective checking if-condition
that does nothing in case the condition is not fulfilled. On the
contrary, by imposing those \ebkeyw{Given} conditions in the US and
eventually in their respective Event-B models I adopt a
\emph{defensive} style of modeling in which the system is required to
be at the right state in order to be able to transition to another
state.

\subsection{Local Invariants for WhatsApp's Basic Functionality}
\label{sub:basic:inv}

When modelling a system in Event-B in addition to the machine's core
functionality, one should write a series of \emph{safety invariant
  properties} that describe the desirable behaviour of the
system. Table \ref{tab:inv:machine0} in Page
\pageref{tab:inv:machine0} presents all the safety invariants that I 
have elicited for WhatsApp's abstract machine.

\subsection{Basic Functionality with Implementation Details}
\label{sub:impdet}

EX-02 offers a general description for the functionality for reading a
chat session. Chat content is read in an orderly fashion.

\begin{center}
\begin{tabular}{|p{1.8cm}|p{9cm}|} \hline
EX-02 & reading-chat \\ \hline
Description & As a user, I want to read a chat session  \\ \hline
Acceptance &\ebkeyw{Given}: A chat session between Me and Another-User exists\\
Criterion  &\ebkeyw{When}: I read a chat session between Me and Another-User \\
  &\ebkeyw{Then}: The content associated to the chat session between Me and Another-user is made available to Me \\ \hline
\end{tabular}
\end{center}
~\\

\section{Basic Functionality of WhatsApp in Event-B}
\label{machine0}

We start by looking at the context of the abstract machine, which
introduces two carrier sets, namely, \ebcons{USER} and \ebcons{CONTENT} that are used to
typeset all the users registered in WhatsApp and all the content that
it manipulates.\\

\begin{tabular}{l}
  \ebkeyw{context} \ebcode{ctx0} \\
  \quad \ebkeyw{sets} \ebcons{USER} \ebcons{CONTENT}\\
  \ebkeyw{end}\\
\end{tabular}
~\\

\begin{tabular}{l}
\ebkeyw{machine} \ebcode{machine0} \ebkeyw{sees} \ebcode{ctx0}\\
~~\ebkeyw{variables} \ebcode{user content chat active chatcontent muted}\\
~\ebcmt{// machine invariants...}\\
\end{tabular}
~\\

\begin{tabular}{l}
~\ebkeyw{event} \ebkeyw{initialisation} \\
~\ebkeyw{then}\\
~~\ebtag{@init1} \ebcode{user :=}  $\emptyset$  ~~~\ebtag{@init2} \ebcode{content :=} $\emptyset$\\
~~\ebtag{@init3} \ebcode{chat :=} $\emptyset$ ~~~\ebtag{@init4} \ebcode{active :=} $\emptyset$\\
~~\ebtag{@init5} \ebcode{chatcontent :=} $\emptyset$ ~~~\ebtag{@init6} \ebcode{muted :=} $\emptyset$ \\
~\ebkeyw{end}\\
~\ebcmt{// rest of machine events...}\\
\ebkeyw{end}
\end{tabular}
~\\

Two variables in our model implement the two first invariants in Table
\ref{tab:inv:machine0}; the first variable stores the registered users and the
second one the content exchanged.  \\

\begin{tabular}{l}
\ebkeyw{invariants}\\
~\ebtag{@inv1} \ebcode{user} $\subseteq$ \ebcons{USER}\\
~\ebtag{@inv2} \ebcode{content} $\subseteq$ \ebcons{CONTENT}\\
\end{tabular}
~\\

Invariant 3 in Table \ref{tab:inv:machine0} is implemented as an Event-B invariant that
declares \ebcode{chat} as a relation between users. Invariant 4 is implemented by the
fact that \ebcode{chat} is a binary relation. Having modelled \ebcode{chat} as a set
enforces the fifth invariant in Table \ref{tab:inv:machine0},
therefore, no pair of elements in a chat session is repeated. \\

\begin{tabular}{l}
~\ebtag{@inv3} \ebcode{chat} $\in$ \ebcode{user} $\rel$ \ebcode{user} \ebcmt{// chat sessions}
\end{tabular}
~\\

Implementing invariant 6 in Table\ref{tab:inv:machine0} requires a subtler
analysis as it relates \ebcode{content}, the sender and the receiver of the
content. Variable \ebcode{chatcontent} below introduces chat content. The variable is
defined as a partial function with domain \ebcode{user} (the person who sends the
message) and range \ebcode{content} $\pfun$ $\pow$\ebcode{(user)}, where
\ebcode{content} is the content sent and $\pow$\ebcode{(user)} is the set of users to
whom the content has been sent. \ebcode{chatcontent} is a partial function, therefore,
it might be the case a user exists that has not chatted with any one. The range of
\ebcode{chatcontent} is again a partial function, therefore, it might be the case a user
exists that has not chatted with some particular user. Since \ebcode{chatcontent} and
its range are functions, the set of users to whom user \ebcode{u1} has sent some content
\ebcode{c} is uniquely represented as \ebcode{chatcontent(u1)(c)}, given that
\ebcode{u1} exists in the domain of \ebcode{chatcontent} and \ebcode{c} exists in the
domain of \ebcode{chatcontent(u1)}. The set of users with whom \ebcode{u1} has chatted
is represented as \ebkeyw{ran}\ebcode{(chatcontent(u1))}, and the set of content items
sent by \ebcode{u1} (to anyone) is represented as \ebkeyw{dom}\ebcode{(chatcontent(u1))},
given that  \ebcode{u1} exists in the domain of \ebcode{chatcontent}.\\

\begin{tabular}{l}
\ebtag{@inv4} \ebcode{chatcontent} $\in$ \ebcode{user} $\pfun$
\ebcode{(content} $\pfun$ $\pow$\ebcode{(user))}
\label{inv:04}
\end{tabular}
~\\

Next, we proceed to encode invariant 8 in Table \ref{tab:inv:machine0} which says that
active and muted chats are disjoints. \ebtag{@inv5} encodes the set of active
chat sessions; \ebcode{active} is a partial function, hence, a user has one active chat session
maximum (the ``function'' part), but it might be the case he has no active chat session at all
(the ``partial'' part). \ebtag{@inv7} states that it is never that case an active chat session is
not a chat session, and \ebtag{@inv8} states that it is never the case that a muted chat session
is not a chat session, that is, elements from muted chats are taken from chats. \ebtag{@inv9}
encodes invariant 8 in Table \ref{tab:inv:machine0}. \\

\begin{tabular}{l}
\ebtag{@inv5} \ebcode{active} $\in$ \ebcode{user} $\pfun$ \ebcode{user} \ebcmt{// active chat session}\\
\ebtag{@inv6} \ebcode{muted} $\in$ \ebcode{user} $\rel$ \ebcode{user} \ebcmt{// muted sessions}\\
\ebtag{@inv7} \ebcode{active} $\subseteq$ \ebcode{chat} \ebcmt{// active chat sessions}\\
\ebtag{@inv8} \ebcode{muted} $\subseteq$ \ebcode{chat} \ebcmt{// muted chat sessions}\\
\ebtag{@inv9} \ebcode{muted} $\cap$ \ebcode{active =} $\emptyset$\\
\end{tabular}
~\\

Invariants 9 and 10 in Table \ref{tab:inv:machine0} state that chat
and active sessions are not symmetric necessarily. This invariants are
modelled by not imposing further constraints over \ebcode{chat} and
\ebcode{active}. In other words, if we wanted them to be symmetric, we
have needed to enforce further invariants in Event-B. Invariant 11 in
Table \ref{tab:inv:machine0} is implemented by \ebtag{@inv10} below. Expression
\ebcode{chat[\{u\}]} returns the set of users with whom user \ebcode{u} is chatting.\\

\begin{tabular}{l}
\ebtag{@inv10} $\forall$\ebcode{u,c,s}$\cdot$\ebcode{u$\in$user $\wedge$ c$\in$content $\limp$}\\
$\quad\quad\quad\quad$\ebcode{(u $\mapsto$ {c $\mapsto$ s} $\in$ chatcontent $\limp$ s $\subseteq$ chat[\{u\}])}\\
\end{tabular}
~\\

Invariant 12 in Table \ref{tab:inv:machine0} is enforced by the fact
that \ebcode{active} is a function.

Event-B models are composed of a static part defining observations (variables,
constants, parameters, etc.) of the system and their invariants properties, and a
dynamic part defining operations (events) changing the state of the system. Definitions
introduced up to now are all static, and the next definitions are the dynamic part of
the abstract machine (\ebcode{machine0)} of our model. Invariant 13 in Table
\ref{tab:inv:machine0} is dynamic. It requires us to add an event guard to every event
that otherwise might modify \ebcode{chatcontent} of a muted chat.

Next, we implement the basic functionality of chat sessions in Event-B. Event
\ebcode{create-chat-session} implements US-01. It creates a chat session for
user \ebcode{u1} to chat with user \ebcode{u2}.  The Given condition in US-01 is
encoded by guard \ebtag{@grd2}. Guard \ebtag{@grd1} 
helps Rodin to infer the type of \ebcode{u1} and \ebcode{u2}. \ebtag{@act1}
adds the pair \ebcode{u1 $\mapsto$ u2} to the set of existing chats. \ebtag{@act2}
makes the content associated to the chat between \ebcode{u1} and \ebcode{u2}
empty. Notice that event \ebcode{create-chat-session} does not create a chat for
\ebcode{u2} to chat with user
\ebcode{u1}.\\

\begin{tabular}{l}
\ebkeyw{event} \ebcode{create-chat-session} \ebcmt{// US-01}\\
\ebkeyw{any} \ebcode{u1 u2}\\
\ebkeyw{where}\\
~\ebtag{@grd1} \ebcode{u1}$\in$\ebcode{user} $\wedge$ \ebcode{u2}$\in$\ebcode{user}\\
~\ebtag{@grd2} \ebcode{u1}$\mapsto$\ebcode{u2} $\notin$ \ebcode{chat}\\
\ebkeyw{then}\\
~\ebtag{@act1} \ebcode{chat := chat $\cup$ \{u1$\mapsto$u2\}}\\
~\ebtag{@act2} \ebcode{active(u1) := u2}\\
\ebkeyw{end}
\end{tabular}
~\\

Event \ebcode{select-chat} implements US-02. \ebtag{@act1} uses the
relational overriding operator $\oplus$ instead of the set union
operator $\cup$, in this way \ebcode{u1} can have an active chat
session only with one user. \ebtag{@grd4} implements a defensive style
of programming as explained before. \ebtag{@grd3} makes sure that a
muted chat session is never active. \ebtag{@grd1} typesets \ebtag{u1}
and \ebcode{u2}. \ebtag{@grd2} implements the first \ebkeyw{Given}
condition in US-02, and guard \ebtag{@grd4} implements the second
one. \ebtag{@act1} uses the overriding operator $\oplus$ instead of
the union operator $\cup$ to make sure we don't break \ebtag{@inv5} so
that \ebcode{active} remains a function. Had we added \ebcode{u1$\mapsto$u2} to
\ebcode{active} using the union operator $\cup$, we would have
probably ended up with \ebcode{active} mapping \ebcode{u1} to two
different users. Rodin would have detected this mistake by generating
an improvable
Proof Obligation (PO).\\

\begin{tabular}{l}
\ebkeyw{event} \ebcode{select-chat} \ebcmt{// US-02}\\
\ebkeyw{any} \ebcode{u1 u2}	\\
\ebkeyw{where}\\
~\ebtag{@grd1} \ebcode{u1$\in$user $\wedge$ u2$\in$user}\\
~\ebtag{@grd2} \ebcode{u1 $\mapsto$ u2 $\in$ chat}\\
~\ebtag{@grd3} \ebcode{u1 $\mapsto$ u2 $\notin$ muted}\\
~\ebtag{@grd4} \ebcode{u1 $\mapsto$ u2 $\notin$ active}\\
\ebkeyw{then}\\
~\ebtag{@act1} \ebcode{active := active $\oplus$ \{u1$\mapsto$u2\}}\\
\ebkeyw{end}\\
\end{tabular}
~\\

Event \ebcode{chatting} implements US-03 whereby user \ebcode{u1}
chats with user \ebcode{u2}. It implements the scenario whereby
\ebcode{u1} sends some content \ebcode{c} to
\ebcode{u2}. \ebtag{@grd2} encodes the \ebkeyw{Given} condition. The
first part of guard \ebtag{@grd4} typesets variable \ebcode{c} and the
second part requires it to be a fresh content. Because \ebcode{c} is a
fresh content, \ebtag{@act1} adds it to the set of
contents. \ebtag{@act2} creates a chat instance for
\ebcode{u2 $\mapsto$ u1} in case it does not exist already. If it
exists, \ebcode{chat} remains unchanged as it is a set. This matches
the actual behaviour of WhatsApp in which a chat window is created for
\ebcode{u2} the first time a user \ebcode{u1} sends her
some content. The second line in \ebtag{@act3} adds \ebcode{c} to
the existing chat content between \ebcode{u1} and
\ebcode{u2}. Notice that \ebcode{chatcontent(u1)} remains a
function after the assignment in \ebtag{@act3} since \ebcode{c} is
not in its domain. \\


\begin{tabular}{l}
\ebkeyw{event} \ebcode{chatting} \ebcmt{// US-03}\\
\ebkeyw{any} \ebcode{u1 u2 c}\\
\ebkeyw{where}\\
~\ebtag{@grd1} \ebcode{u1$\in$user $\wedge$ u2$\in$user}\\
~\ebtag{@grd2} \ebcode{u1 $\mapsto$ u2 $\in$ active}\\
~\ebtag{@grd3} \ebcode{u1 $\mapsto$ u2 $\notin$ muted $\wedge$ u2
$\mapsto$ u1 $\notin$ muted}\\
~\ebtag{@grd4} \ebcode{c $\in$} \ebcons{CONTENT} $\wedge$ \ebcode{c $\notin$ content} \\
~\ebtag{@grd5} \ebcode{u1 $\in$ \ebkeyw{dom}(chatcontent)} \\
\ebkeyw{then}\\
~\ebtag{@act1} \ebcode{content := content} $\cup$ \ebcode{\{c\}}\\
~\ebtag{@act2} \ebcode{chat := chat $\cup$ \{u2 $\mapsto$ u1\}}\\
~\ebtag{@act3} \ebcode{chatcontent := chatcontent $\oplus$} \\
\quad\quad\quad\quad\quad\quad\quad\quad\quad\ebcode{\{ u1 $\mapsto$
(chatcontent(u1) $\cup$ \{c~$\mapsto$~\{u2\}\})~\}}\\
\ebkeyw{end}\\
\end{tabular}
~\\

We present below the encoding of US-04a and US-04b, therefore, guard
\ebtag{@grd2} verifies that the user \ebcode{u1} who deletes or
removes the content is actively chatting with
\ebcode{u2}. \ebcode{delete-content} uses the functional overriding
operator $\oplus$ to override \ebcode{u1}'s chat content. It removes
\ebcode{u2} from \ebcode{chatcontent(u1)(c)} so that \ebcode{u2} no
longer appears as having received content \ebcode{c} from \ebcode{u1}.\\

\begin{tabular}{l}
\ebkeyw{event} \ebcode{delete-content} \ebcmt{// US-04a}\\
\ebkeyw{any} \ebcode{u1 u2 c}\\
\ebkeyw{where}\\
~\ebtag{@grd1} \ebcode{u1$\in$user $\wedge$ u2$\in$user}\\
~\ebtag{@grd2} \ebcode{u1$\mapsto$u2 $\in$ active}\\
~\ebtag{@grd3} \ebcode{u1 $\in$ dom(chatcontent)}\\
~\ebtag{@grd4} \ebcode{c $\in$ dom(chatcontent(u1))}\\
~\ebtag{@grd5} \ebcode{u2 $\in$ chatcontent(u1)(c)}\\
\ebkeyw{then}\\
~\ebtag{@act1} \ebcode{chatcontent(u1) := chatcontent(u1)} $\oplus$ \\
\quad\quad\quad\quad\quad\quad\quad\quad\quad\quad\quad\ebcode{\{c $\mapsto$ (chatcontent(u1)(c)$\setminus$\{u2\})\}}\\
\ebkeyw{end}\\
\end{tabular}
~\\

\ebcode{remove-content} removes \ebcode{c} from the domain of
\ebcode{chatcontent(u1)}, therefore,
\ebcode{c} no longer appears as having been sent by \ebcode{u1}. \\

\begin{tabular}{l}
\ebkeyw{event} \ebcode{remove-content} \ebcmt{// US-04b}\\
\ebkeyw{any} \ebcode{u1 u2 c}\\
\ebkeyw{where}\\
~\ebtag{@grd1} \ebcode{u1$\in$user $\wedge$ u2$\in$user}\\
~\ebtag{@grd2} \ebcode{u1$\mapsto$u2 $\in$ active}\\
~\ebtag{@grd3} \ebcode{u1 $\in$ dom(chatcontent)}\\
~\ebtag{@grd4} \ebcode{c $\in$ dom(chatcontent(u1))}\\
~\ebtag{@grd5} \ebcode{u2 $\in$ chatcontent(u1)(c)}\\
\ebkeyw{then}\\
~\ebtag{@act1} \ebcode{chatcontent(u1) := \{c\} $\domsub$ chatcontent(u1)}\\
~\ebtag{@act2} \ebcode{content := content$\setminus$\{c\}}\\
\ebkeyw{end}\\
\end{tabular}
~\\

Event \ebcode{mute-chat} encodes US-06. It mutes the chat between \ebcode{u1}
and \ebcode{u2}; more concretely \ebtag{@act1} adds the pair
\ebcode{u1$\mapsto$u2} to the set of muted chats. \ebtag{@act2} forbids a muted chat
from being active. Alternatively, we could have added a guard \ebtag{@grd4} \ebcode{u1 $\mapsto$
u2 $\notin$ active}, but then this does not reflect the actual behaviour of the
graphical interface of WhatsApp in which a user \ebcode{u1} can indeed mute a user \ebcode{u2}
when actively chatting with her.\\

\begin{tabular}{l}
\ebkeyw{event} \ebcode{mute-chat} \ebcmt{// US-06}\\
\ebkeyw{any} \ebcode{u1 u2}\\
\ebkeyw{where}\\
~\ebtag{@grd1} \ebcode{u1$\in$user $\wedge$ u2$\in$user}\\
~\ebtag{@grd2} \ebcode{u1 $\mapsto$ u2 $\in$ chat}\\
~\ebtag{@grd3} \ebcode{u1 $\mapsto$ u2 $\notin$ muted}\\
\ebkeyw{then}\\
~\ebtag{@act1} \ebcode{muted := muted $\cup$ \{u1$\mapsto$u2\}}\\
~\ebtag{@act2} \ebcode{active := active $\setminus$ \{u1$\mapsto$u2\}}\\
\ebkeyw{end} \\
\end{tabular}
~\\

Event \ebcode{unmute-chat} implements US-07. It unmutes the chat
between \ebcode{u1} and \ebcode{u2}. \ebtag{@grd3} checks that user
\ebcode{u1} (who mutted \ebcode{u2}) is the only one who can unmmute
\ebcode{u2}. \ebcode{u2} is unique since \ebcode{muted} is a
function. \ebtag{@grd2} is redundant: it can be deduced from
\ebtag{@grd3} and the fact that \ebcode{muted $\subseteq$
  chat}. 
\\

\begin{tabular}{l}
\ebkeyw{event} \ebcode{unmute-chat} \ebcmt{// US-07}\\
\ebkeyw{any} \ebcode{u1 u2}\\
\ebkeyw{where}\\
~\ebtag{@grd1} \ebcode{u1$\in$user $\wedge$ u2$\in$user}\\
~\ebtag{@grd2} \ebcode{u1 $\mapsto$ u2 $\in$ chat}\\
~\ebtag{@grd3} \ebcode{u1 $\mapsto$ u2 $\in$ muted}\\
\ebkeyw{then}\\
~\ebtag{@act1} \ebcode{muted := muted $\setminus$ \{u1$\mapsto$u2\}}\\
\ebkeyw{end}\\
\end{tabular}
~\\

Event \ebcode{forward} below implements US-09 whereby user \ebcode{u}
forwards content \ebcode{c} to a set of users \ebcode{us}. Guards
\ebtag{@grd1}, \ebtag{@grd2}, and \ebtag{@grd5} typeset \ebcode{u} and
\ebcode{us}. \ebtag{@grd4} typesets \ebcode{c}. \ebtag{@grd6} checks
that \ebcode{u} indeed possesses chat sessions with every user member
of \ebcode{us}. Expression \ebcode{muted[\{u\}]} $\cap$ \ebcode{us =}
$\emptyset$ checks that no member of the set \ebcode{us} is part of
the set of users that \ebcode{u} has muted. Expression
\ebcode{muted[us] $\cap$ \{u\} =} $\emptyset$ checks that \ebcode{u}
has not been muted by any member of \ebcode{us}. Body action
\ebtag{@act2} creates respective chat sessions for each member of the
set \ebcode{us} to chat with \ebcode{u}. Action \ebtag{@act1}
overrides \ebcode{chatcontent} to include content item \ebcode{c} into
the chat sessions between \ebcode{u} and each element of
\ebcode{us}. Event \ebcode{forward} itself does not encode a notion of
order among the content items of a chat session, hence, at the
abstract level as implemented by the abstract machine one cannot
establish which chat content reads first or after another.  Event
\ebcode{broadcast} for US-08 is implemented in a similar way to event
\ebcode{forward}. The major difference between their implementations
is that guard \ebtag{@grd6} for event \ebcode{forward} is not included
by event \ebcode{broadcast}.\\

\begin{tabular}{l}
\ebkeyw{event} \ebcode{forward} \ebcmt{// US-09}\\
\ebkeyw{any} \ebcode{u us c}\\
\ebkeyw{where}\\
~\ebtag{@grd1} \ebcode{u $\in$ user}\\
~\ebtag{@grd2} \ebcode{us $\subseteq$ user}\\
~\ebtag{@grd3} \ebcode{muted[\{u\}] $\cap$ us = $\emptyset$ $\wedge$ muted[us] $\cap$ \{u\} = $\emptyset$}\\
~\ebtag{@grd4} \ebcode{c $\in$ content}\\
~\ebtag{@grd5} \ebcode{u $\in$ dom(chatcontent)}\\
~\ebtag{@grd6} \ebcode{us $\subseteq$ chat[\{u\}]}\\
\ebkeyw{then}\\
~\ebtag{@act1} \ebcode{chatcontent := chatcontent $\oplus$ \{u $\mapsto$ (chatcontent(u) $\cup$ \{c $\mapsto$ us\})\}}\\
~\ebtag{@act2} \ebcode{chat := chat $\cup$ (us $\times$ \{u\})}\\
\ebkeyw{end}\\
\end{tabular}
~\\

Event \ebcode{unselect-chat} implements US-10. It unselects the chat between
\ebcode{u1} and \ebcode{u2} by dropping \ebcode{u1 $\mapsto$ u2} from
\ebcode{active}.\\

\begin{tabular}{l}
\ebkeyw{event} \ebcode{unselect-chat} \ebcmt{// US-10}\\
\ebkeyw{any} \ebcode{u1 u2}\\
\ebkeyw{where}\\
~\ebtag{@grd1} \ebcode{u1$\in$user $\wedge$ u2$\in$user}\\
~\ebtag{@grd2} \ebcode{u1 $\mapsto$ u2 $\in$ chat}\\
~\ebtag{@grd3} \ebcode{u1 $\mapsto$ u2 $\in$ active}\\
\ebkeyw{then}\\
~\ebtag{@act1} \ebcode{active := active $\setminus$ \{u1$\mapsto$u2\}}\\
\ebkeyw{end}\\
\end{tabular}
~\\

\section{Design and Implementation Decisions Regarding \ebcode{machine0}}
\label{decisions0}

Event \ebcode{create-chat-session}. What are the consequences of
making \ebcode{a1 $\mapsto$ a2} \ebcode{active}? Rodin discharge its
POs automatically, hence \ebcode{create-chat-session} is correct with
respect to the invariants defined in \ebcode{machine0}. The
consequences of making or not \ebcode{a1 $\mapsto$ a2} \ebcode{active}
are rather related with its inter-playing with other events, for
instance, with events \ebcode{chatting} and \ebcode{select-chat}. If
\ebcode{create-chat-session} doesn't make \ebcode{a1 $\mapsto$ a2}
\ebcode{active} then \ebcode{select-chat} should execute later on
before start chatting. The analysis of the inter-playing in the
execution of several events in an Event-B model can typically be
performed using ProB~\cite{prob}. This tool checks for deadlocks. It
checks if after executing any event the system can always make
progress or not.

Event \ebcode{chatting}. The first decision to make is whether or not
we want to add content \ebcode{c} to the chat between \ebcode{u2} and
\ebcode{u1} (in addition to the chat between \ebcode{u1} and
\ebcode{u2}). If we want to do so, we should extend the second line of
\ebtag{@act3} with \ebcode{u2 $\mapsto$ (chatcontent(u2) $\cup$
  \{c~$\mapsto$~\{u1\}\})}. Intuitively, adding this line means that
the content \ebcode{c} that \ebcode{u1} sends to \ebcode{u2} is not
only seen by \ebcode{u1} but also by \ebcode{u2}. However, if we
choose to extend \ebtag{@act3} that way, Rodin provers would generate
a PO henceforth \ebcode{u2} must be in the
\ebcode{\ebkeyw{dom}(chatcontent)} so that sub-expression
\ebcode{chatcontent(u2)} is well-typed. This requirement can be solved
by adding an event guard \ebtag{@grd6} \ebcode{u2 $\in$
  \ebkeyw{dom}(chatcontent)}. The downside of this solution is that
\ebtag{@grd6} does not hold the first time when \ebcode{u2} hasn't
sent any content to anyone (not just to \ebcode{u1}) yet. In other
words, the first time that \ebcode{u1} chats with \ebcode{u2} no 
chat session \ebcode{u2 $\mapsto$ u1} exists yet.

The above downside would suggest that one could add default chat
content associations the first time that one creates \ebcode{u2} (or
any user, in general). The event \ebcode{add-user} below adds user
\ebcode{u} to the set of current users. Action \ebtag{@act2} adds
default chat content associations for user \ebcode{u} with respect to
any existing \ebcode{content}. The soundness of \ebtag{@act2} is
corroborated by Rodin provers by discharging all the associated POs
automatically; in particular, \ebtag{@act2} adheres to \ebtag{@inv4}
in Page \pageref{inv:04}.\\

\begin{tabular}{l}
\ebkeyw{event} \ebcode{add-user}\\
\ebkeyw{any} \ebcode{u}\\
\ebkeyw{where}\\
 ~\ebtag{@grd1} \ebcode{u $\in$ \ebcons{USER} $\setminus$ user}\\
\ebkeyw{then}\\
 ~\ebtag{@act1} \ebcode{user := user $\cup$ \{u\}}\\
 ~\ebtag{@act2} \ebcode{chatcontent(u) := content $\times$ \{$\emptyset$\}}\\
\ebkeyw{end}\\
\end{tabular}
~\\

What would it happen with the association encoded by \ebtag{@act2}
above the next time that we add (create) a new content item? Event
\ebcode{add-content} is shown below. For each and every existing
\ebcode{user}, \ebtag{@act2} associates the fresh content \ebcode{c}
to the empty set, in other words, content item \ebcode{c} appears as
been sent by the whole set of users \ebcode{user} to anyone.\\

\begin{tabular}{l}
\ebkeyw{event} \ebcode{add-content}\\
\ebkeyw{any} \ebcode{c}\\
\ebkeyw{where}\\
 ~\ebtag{@grd1} \ebcode{c $\in$ \ebcons{CONTENT} $\setminus$ content}\\
\ebkeyw{then}\\
 ~\ebtag{@act1} \ebcode{content := content $\cup$ \{c\}}\\
 ~\ebtag{@act2} \ebcode{chatcontent := chatcontent $\cup$
(user$\times$\{\{c $\mapsto \emptyset$\}\})}\\
\ebkeyw{end}\\
\end{tabular}
~\\

Summing up on event \ebcode{chatting}, if we wanted to add content
item \ebcode{c} to chat \ebcode{u2 $\mapsto$ u1} in addition to chat
{u1 $\mapsto$ u2}, then we would incur into a computationally
expensive task: we would need to associate $\emptyset$ to every
existing content item every time we add a user to the system, and we
would need to associate every single user to \ebcode{\{c $\mapsto
 \emptyset$\}} every time we needed to add a fresh content item
\ebcode{c}. This type of analysis on the complexity of associating
chat content to \ebcode{u2 $\mapsto$ u1}  is not very intricate, in
general; this analysis can be performed through careful code
inspection or testing. But, writing the formal specification of
WhatsApp in Event-B forces one to do code-inspection, and having Rodin
theorem provers ensures that all cases are considered when performing
automatic checking of Event-B specifications with Rodin, without
having to put effort into writing appropriate test scenarios. 

Notice that expressing \ebtag{@act3} in \ebcode{chatting} as below
does not work since the last overriding expression forgets about
\ebcode{chatcontent(u2)}, which amounts to deleting it. This issue
cannot be spotted by Rodin (in particular regarding invariant
\ebtag{@inv4}) as the new association for \ebcode{u2} would
still be a partial function. This can only be spotted by a domain
expert who knows that she does not want her chat to be
deleted whenever a content is sent to her.\\

\begin{tabular}{l}
\ebtag{@act3} \ebcode{chatcontent := chatcontent}\\
\quad\quad\ebcode{$\oplus$ \{u1 $\mapsto$ (chatcontent(u1) $\cup$ \{c $\mapsto$ \{u2\}\})}\\
\quad\quad\ebcode{$\oplus$ \{u2 $\mapsto$ \{c $\mapsto$ \{u1\}\}\}}\\
\end{tabular}
~\\

The final solution is to use a comprehension set expression to express the new
value of \ebcode{chatcontent(u2)} as indicated in the last overriding expression
below. The downside of this solution is that this expression is not directly
encoded with sets, relations and their operators (domain restriction, domain
subtraction, inverse, etc.), which are, for instance, directly encoded into Java
by Event-B code generators like EventB2Java
\cite{b2jml:tool:16,B2Jml:12,sttt:codegen:17}. \\

\begin{tabular}{l}
\ebtag{@act3} \ebcode{chatcontent := chatcontent}\\
\quad\quad \quad\quad\ebcode{$\oplus$ \{u1 $\mapsto$ (chatcontent(u1) $\cup$ \{c $\mapsto$ \{u2\}\})\}}\\
\quad\quad\quad\quad\ebcode{$\oplus$ \{cc,s $\cdot$ u2 $\mapsto$ \{cc $\mapsto$ s\} $\in$
  chatcontent $\vee$ (cc=c $\wedge$   s=\{u1\})}\\
\quad\quad\quad\quad\quad\quad\quad$\;\;$\ebcode{$|$ u2 $\mapsto$ \{cc $\mapsto$ s\}\}}\\
\end{tabular}
~\\

\ebcode{delete-content} and \ebcode{remove-content} are two of the
subtlest functionality of WhatsApp in the sense that performing either
of them can potentially break invariants all around. Notice that
\ebcode{delete-content} does not remove c from chat
\ebcode{u2 $\mapsto$ u1}. Under which circumstances should one add
the following action to event \ebcode{delete-content}? \\

\begin{tabular}{l}
\ebtag{@act2} \ebcode{content := content $\setminus$ \{c\}}\\
\end{tabular}
~\\

If we add \ebtag{@act2} to \ebcode{delete-content}, Rodin will
generate an unprovable PO. The PO is related to \ebtag{@inv4} in Page
\pageref{inv:04}. One would need to demonstrate that for any user
\ebcode{u} other than \ebcode{u1} the range of \ebcode{chatcontent(u)}
is a partial function from \ebcode{content$\setminus$\{c\}} to
$\pow$\ebcode{(user)}, which is not possible because it might be the case
that \ebcode{u} has sent (forwarded or broadcasted) \ebcode{c} to
another user previously. 

A turn-around to this problem is to express \ebcode{content} as below. However, to
calculate the value of \ebcode{chatcontent} that way one should traverse
\ebcode{user} twice and \ebcode{content}) once, which might be time consuming
depending on the type of structures used to store \ebcode{chatcontent} or to represent sets in general. \\

\begin{tabular}{l}
\ebtag{@act2} \ebcode{content := \{cc,a,b,s $\cdot$ a$\in$\ebkeyw{dom}(chatcontent) $\wedge$}  \\
\quad\quad\ebcode{cc $\mapsto$ s $\in$ chatcontent(a) $\wedge$ b$\in$s
 $\wedge$ $\neg$(a=u1 $\wedge$ b=u2 $\wedge$ cc=c) $|$ cc\}}
\end{tabular}
~\\

Notice that if \ebcode{u1} is chatting with \ebcode{u2}, and
\ebcode{u2} with \ebcode{u3}, and \ebcode{u1} sends \ebcode{c} to
\ebcode{u2}, and \ebcode{u2} sends \ebcode{c} to \ebcode{u3}, calling
\ebcode{remove-content} with parameters \ebcode{u1}, \ebcode{u2}, and
\ebcode{c} does not remove \ebcode{c} from the chat between
\ebcode{u2} and \ebcode{u3}, but only from the chat between
\ebcode{u1} and \ebcode{u2} and between \ebcode{u2} and
\ebcode{u1}. For this reason \ebcode{remove-content} does not
implement a second action \ebtag{@act2} \ebcode{content :=
 content$\setminus$\{c\}}. To express the new value of
\ebcode{content} we can adopt the same approach as above and add
the following line to the event \ebcode{remove-content}.\\

\begin{tabular}{l}
\ebtag{@act2} \ebcode{content := \{cc,a,s $\cdot$ a$\in$\ebkeyw{dom}(chatcontent) $\wedge$}  \\
\quad\quad\ebcode{cc $\mapsto$ s $\in$ chatcontent(a) 
 $\wedge$ $\neg$(a=u1 $\wedge$ cc=c) $|$ cc\}}
\end{tabular}
~\\

Event \ebcode{forward}. It presents the same problem as event
\ebcode{chatting}. That is, if we additionally want to augment
\ebcode{chatcontent} with triplets \ebcode{u2 $\mapsto$ \{c $\mapsto$ \{u\}\}}
for each \ebcode{u2$\in$us}, then we would need to add a second overriding
expression like the one shown below. Again, set comprehension expressions are
not implemented by tools like the EventB2Java Java code generator and
hence that expression would need to be (machine) refined before it can be
translated to a language like Java.\\

\begin{tabular}{l}
\ebtag{@act1} \ebcode{chatcontent := chatcontent}\\
\quad\ebcode{$\oplus$ \{u $\mapsto$ (chatcontent(u) $\cup$ \{c $\mapsto$ us\})\}}\\
\quad\ebcode{$\oplus$ \{u2,cc,s $\cdot$ (u2 $\mapsto$ \{cc $\mapsto$ s\}$\in$chatcontent) $\vee$ (cc=c $\wedge$ s=\{u\} $\wedge$  u2$\in$us)} \\
\quad\quad\quad\quad\quad\quad\ebcode{$|$ u2 $\mapsto$ \{cc $\mapsto$ s\}\}}
\end{tabular}
~\\

Event \ebcode{mute-chat}. It encodes a \emph{defensive} style of programming
whereby only unmuted chats can then be muted. If the chat \ebcode{u1 $\mapsto$
  u2} is muted, then event \ebcode{mute-chat} does not execute. However, notice
that if we were to execute \ebtag{@act1} with a muted chat, then state variable
\ebcode{muted} would be remain unchanged as sets do not contain repeated
elements (see \ebtag{@act1}).


\section{Extended Functionality with Implementation Details}
\label{machine2}
\emph{Machine refinement} is the mechanism that Event-B offers to
extend or to detail the behaviour and the functionality of a
machine. In Event-B, all the components of a refined, machine variable
initialisations, guards and actions of a refining event defined using
refines) or implicitly (invariants, guards and actions of a refining
event defined using extends). We don't give details here about
\ebcode{machine1} (the first machine refinement) but rather focus on
the encoding of \ebcode{machine2}. This machine adds implementation
details to our Event-B model of WhatsApp. The goal of this machine is
to leave the Event-B model into a way that is close to implementation
for it to be translated to Java using the EventB2Java tool. Variables
of a refined machine can appear in an invariant of a refinement
machine. When this happens, the invariant is called a \emph{gluing
  invariant} as it relates the state space of the abstract (refined)
machine with the state space of the refinement machine. Until now, we
have worked content, users, and chat content with an abstract data
structure \emph{set}. This structure was chosen for clarity rather
than for its ability to be implemented in a computer. In their
implementation in \ebcode{machine2} we want to represent (some of)
these structures with \emph{sequences}.

In Event-B, a segment of natural numbers can be expressed using the 
\ebcode{a..b} notation, which defines the set of natural numbers between
\ebcode{a} and \ebcode{b} inclusive.\\

\ebcode{a..b = \{x $|$ x$\in\nat$ $\wedge$ a$\leq$x $\wedge$ x$\leq$b\}}\\

We can hence use the \ebcode{1..n} notation to model a sequence of
type \ebcode{T} and size \ebcode{n} as a total function from
\ebcode{1..n} to \ebcode{T}. By requiring \ebcode{sequence} to be a
total function we enforce it to have no holes in its domain. \\

\ebtag{@inv} \ebcode{sequence $\in$ 1..n $\tfun$ T}\\

Variable \ebcode{contents} below encodes \ebcode{content} as a
sequence. \ebcode{csize} represents the number of content items in
\ebcode{contents}. \ebcode{content} is the type of
\ebcode{contents}. The domain of \ebcode{contents} is
\ebcode{1..csize}, hence, when \ebcode{csize} is \ebcode{0},
\ebcode{contents} is empty. \ebtag{@invr22} and \ebtag{@invr23} are
together a {gluing invariant} that relates \ebcode{contents} with 
\ebcode{content}.\\

\begin{tabular}{l}
~\ebtag{@invr21} \ebcode{csize $\ge$ 0}\\
~\ebtag{@invr22} \ebcode{contents $\in$ (1 .. csize) $\tsur$ content}\\
~\ebtag{@invr23} \ebcode{content = \{n,c $\cdot$ n $\mapsto$ c $\in$ contents $|$ c\}}\\
\end{tabular}
~\\

Next, we choose to implement \ebcode{chatcontent} as the variable
\ebcode{screen}. This refined variable makes content-sent sequential,
but not the sender or the receiver of the content. This is because we
mainly use \ebcode{screen} to display content exchanged in an orderly
fashion, for which the pair of users do not need to be ordered, just
the content items. \\

\ebtag{@invr24} \ebcode{screen $\in$ user $\pfun$ (user $\pfun$ $\pow$(contents))}\\





We present the refined version of event \ebcode{chatting}
below. Parameter \ebcode{k1} is the position at which content
\ebcode{c} is placed at \ebcode{u1}'s screen. For \ebcode{u1}'s chat
content to be shown in an orderly fashion, \ebcode{k1} must be greater
than any value in \ebcode{\ebkeyw{dom}(screen(u1)(u2))} every time
that \ebcode{chatting} executes. Likewise, \ebcode{k2} is the position
of content item \ebcode{c} in \ebcode{u2}'s chat screen with
\ebcode{u1}. The last conjuncts in \ebtag{@grdr21} and \ebtag{@grdr22}
together ensure that $\pow$\ebcode{(contents)} in \ebtag{@invr24} is a
function. Action \ebtag{@actr22} increases the number of existing
content items. Action \ebtag{@actr23} adds \ebcode{c} at position
\ebcode{csize+1} of sequence \ebcode{contents}. \ebtag{@actr21} adds
\ebcode{c} at position \ebcode{k1} (\ebcode{k2}) of \ebcode{u1}'s
(\ebcode{u2}'s) chat screen with \ebcode{u2} (\ebcode{u1}). \\

\begin{tabular}{l}
\ebkeyw{event} \ebcode{chatting} \ebkeyw{extends} \ebcode{chatting} \ebcmt{// US-03}\\
\ebkeyw{any} \ebcode{k1 k2}\\
\ebkeyw{where}\\
~\ebtag{@grdr21} \ebcode{u1 $\in$ \ebkeyw{dom}(screen) $\wedge$ u2 $\in$ \ebkeyw{dom}(screen(u1)) $\wedge$}\\
\quad\quad\quad\quad\quad\ebcode{k1 $\not\in$ \ebkeyw{dom}(screen(u1)(u2))}\\ 
~\ebtag{@grdr22} \ebcode{u2 $\in$ \ebkeyw⁄{dom}(screen) $\wedge$ u1 $\in$ \ebkeyw{dom}(screen(u2)) $\wedge$}\\
\quad\quad\quad\quad\quad\ebcode{k2 $\notin$ \ebkeyw{dom}(screen(u2)(u1))}\\
\ebkeyw{then}\\
 ~\ebtag{@actr21} \ebcode{screen := screen $\oplus$} \\
  \quad\quad\quad\quad\quad\ebcode{\{u1 $\mapsto$ (screen(u1) $\oplus$ \{u2 $\mapsto$ (screen(u1)(u2) $\oplus$} \ebcode{\{k1 $\mapsto$ c\})\}),}\\
  \quad$\quad\quad\quad\quad\;\;$\ebcode{u2 $\mapsto$ (screen(u2) $\oplus$ \{u1 $\mapsto$  (screen(u2)(u1) $\oplus$} \ebcode{\{k2 $\mapsto$ c\})\})\}}\\
 ~\ebtag{@actr22} \ebcode{csize := csize+1}\\
 ~\ebtag{@actr23} \ebcode{contents := contents $\oplus$ \{(csize+1) $\mapsto$ c\}}\\
\ebkeyw{end} \\
\end{tabular}
~\\

Event \ebcode{delete-content} declares two parameters \ebcode{i} and
\ebcode{k} for the position of content \ebcode{c} in the sequences
\ebcode{contents} and \ebcode{screen}, respectively. \ebtag{@grdr21}
checks that \ebcode{contents(i) = c}. The last conjunct of
\ebtag{@grdr22} checks that \ebcode{c} is displayed at position
\ebcode{k} of the chat screen between \ebcode{u1} and
\ebcode{u2}. \ebtag{@actr21} deletes \ebcode{k $\mapsto$ c} from
\ebcode{screen(u1)(u2)}. \\

\begin{tabular}{l}
\ebkeyw{event} \ebcode{delete-content \ebkeyw{extends} delete-content} \ebcmt{// US-04}\\
\ebkeyw{any} \ebcode{i k}\\
\ebkeyw{where}\\
~\ebtag{@grdr21} \ebcode{i $\mapsto$ c $\in$ contents}\\
~\ebtag{@grdr22} \ebcode{u1 $\in$ \ebkeyw{dom}(screen) $\wedge$ u2 $\in$ \ebkeyw{dom}(screen(u1)) $\wedge$}\\
\quad\quad\quad\quad\quad\ebcode{k $\mapsto$ c $\in$ screen(u1)(u2)}\\
\ebkeyw{then}\\
~\ebtag{@actr21} \ebcode{screen(u1) := screen(u1) $\oplus$ \{u2 $\mapsto$ (\{k\} $\domsub$ screen(u1)(u2))\}}\\
\ebkeyw{end}\\
\end{tabular}
~\\

\section{Design and Implementation Decisions Regarding \ebcode{machine2}}
\label{decisions2}

Event \ebcode{chatting}. Some of the previous discussions in Section \ref{decisions0} are
revisited in this section. We could opt to define in \ebcode{machine2} an additional
\ebcode{chatting-first-time} event that works the first time that user \ebcode{u1} sends
any content to user \ebcode{u2} (hence \ebcode{u1} doesn't exist in the domain of
\ebcode{screen(u2)}). Therefore, we redefine guard \ebtag{@grdr22} to reflect the case
when \ebcode{u1$\not\in$\ebkeyw{dom}(screen(u2))}, and modify \ebtag{@actr21} not to
refer to \ebcode{screen(u2)(u1)}. \\

\begin{tabular}{l}
\ebkeyw{event} \ebcode{chatting-first-time} \ebkeyw{extends} \ebcode{chatting} \ebcmt{// US-03}\\
\ebkeyw{any} \ebcode{k1 k2}\\
\ebkeyw{where}\\
~\ebtag{@grdr21} \ebcode{u1 $\in$ \ebkeyw{dom}(screen) $\wedge$ u2$\in$\ebkeyw{dom}(screen(u1)) $\wedge$}\\
\quad\quad\quad\quad\quad\ebcode{k1 $\not\in$ \ebkeyw{dom}(screen(u1)(u2))}\\
~\ebtag{@grdr22} \ebcode{u2 $\in$ \ebkeyw{dom}(screen) $\wedge$ u1 $\not\in$ \ebkeyw{dom}(screen(u2)) $\wedge$}\\
\quad\quad\quad\quad\quad\ebcode{k2 $\in$ $\intg$}\\
\ebkeyw{then}\\
 ~\ebtag{@actr21} \ebcode{screen := screen $\oplus$} \\
  \quad\quad\quad\quad\quad\ebcode{\{u1 $\mapsto$ (screen(u1) $\oplus$ \{u2 $\mapsto$ (screen(u1)(u2) $\oplus$} \ebcode{\{k1 $\mapsto$ c\})\}),}\\
  \quad$\quad\quad\quad\quad\;\;$\ebcode{u2 $\mapsto$ (screen(u2) $\oplus$ \{u1 $\mapsto$} \ebcode{\{k2 $\mapsto$ c\}\})\}}\\
 ~\ebtag{@actr22} \ebcode{csize := csize+1}\\
 ~\ebtag{@actr23} \ebcode{contents := contents $\oplus$ \{(csize+1) $\mapsto$ c\}}\\
\ebkeyw{end} \\
\end{tabular}
~\\

If one wants to show the content of a chat in an orderly fashion, then
\ebcode{chatting} should always be fed with indexes \ebcode{k1} and \ebcode{k2} that
are greater than any previous index.

Event \ebcode{delete-content}. Expressing the new value of \ebcode{contents} without
recurring to the use of a set comprehension expression is a difficult problem for
concrete machine \ebcode{machine2} as well. By simply looking at the type given by
\ebtag{@invr24} to \ebcode{screen}, if content \ebcode{c} is deleted from
\ebcode{screen(u1)(u2)}, one would need to search for \ebcode{c} in every element of
type \ebcode{$\pow$(contents)} associated to every pair of users in \ebcode{screen}; if
\ebcode{c} is ever found, then \ebcode{contents} remains unchanged, otherwise,
\ebcode{contents} is modified so that it becomes \ebcode{contents $\setminus$ \{c\}}. This is
an algorithmic solution that can be implemented in a streamline programming language,
which would be difficult to express in logic using sets, relations and operators over them only.


Event \ebcode{forward}. We show below an unsuccessful attempt to implement event 
\ebcode{forward}\footnote{\ebkeyw{card} returns the number of elements of a
  set.}. The event maps \ebcode{c} with a new index
\ebcode{k$\not\in$\ebkeyw{dom}(screen(u)(u2))} for each user \ebcode{u2} in the set
of users \ebcode{us}. However, $(i.)$ \ebcode{k} ought to be the maximum element in
\ebcode{\ebkeyw{dom}(screen(u)(u2))} for each user \ebcode{u2}, and $(ii.)$ content
\ebcode{c} must be seen in \ebcode{screen(u2)(u)} in addition to
\ebcode{screen(u)(u2)}.\\

\begin{tabular}{l}
  \ebkeyw{event} \ebcode{forward} \ebkeyw{extends} \ebcode{forward} \\
  \ebkeyw{any} \ebcode{ks}\\
  \ebkeyw{where}\\
  ~\ebtag{@grdr21} \ebcode{u $\in$ \ebkeyw{dom}(screen) $\wedge$ us $\subseteq$ \ebkeyw{dom}(screen(u))}\\
  ~\ebtag{@grdr22} \ebcode{ks $\subseteq \nat$ $\wedge$ \ebkeyw{card}(ks) = \ebkeyw{card}(us)}\\
  \ebkeyw{then}\\
  ~\ebtag{@actr21} \ebcode{screen := screen $\oplus$}\\
  \quad\quad\quad\ebcode{\{ u $\mapsto$ (screen(u) $\oplus$} \ebcode{\{u2,k $\cdot$ u2$\in$us $\wedge$ k$\not\in$\ebkeyw{dom}(screen(u)(u2))}\\
  \quad\quad\quad\ebcode{$|$ u2 $\mapsto$ (screen(u)(u2) $\oplus$ \{k $\mapsto$ c\})\}) \}}\\    
  \ebkeyw{end}\\
\end{tabular}
~\\

The encoding below addresses issue $(i.)$, hence, as for \ebtag{@grdr22}, the user
needs to feed event \ebcode{forward} with a parameter $k$ that is greater than any
element in \ebcode{screen(u)(u2)}. Code generators like EventB2Java will implement event
guards through if-conditions. However, coding \ebtag{@grdr22} with an if-condition
would negatively affect the performance of the event implementation as the
underlying checking would need to be performed every time the event is to be
executed. One can therefore think of moving the checking \ebtag{@grdr22} outside
the event and entrust it to a function that keeps (or calculates) a maximum
\ebcode{k} value for each pair of users \ebcode{u} and \ebcode{u2} and then calling that
function every time a \ebcode{k} is needed.\\

 \begin{tabular}{l}
   \ebkeyw{event} \ebcode{forward} \ebkeyw{extends} \ebcode{forward} \ebcmt{// US-09} \\
   \ebkeyw{any} \ebcode{k}\\
   \ebkeyw{where}\\
   ~\ebtag{@grdr21} \ebcode{u $\in$ \ebkeyw{dom}(screen) $\wedge$ us $\subseteq$ \ebkeyw{dom}(screen(u))}\\
   ~\ebtag{@grdr22} \ebcode{$\forall$ u2, i $\cdot$ u2$\in$us $\wedge$ i$\in$dom(screen(u)(u2)) $\limp$ k$>$i}\\
   \ebkeyw{then}\\
   ~\ebtag{@actr21} \ebcode{screen := screen $\oplus$ \{ u $\mapsto$ (screen(u) $\oplus$}\\
   \quad\quad\quad\quad\quad\quad\quad\quad\quad\ebcode{\{u2 $\cdot$  u2$\in$us $|$ u2 $\mapsto$ (screen(u)(u2) $\oplus$ \{k$\mapsto$c\})\}) \}}\\
   \ebkeyw{end}\\
 \end{tabular}
 ~\\

 Event \ebcode{broadcast}. The same problems for \ebcode{forward} apply to
 \ebcode{broadcast}.

\section{Conclusion}
\label{sec:con}

Formal Methods (FMs) will become more popular in the Smart Phone
industry if techniques are developed and tools are implemented that
provide support to Software Engineering practices and to the analysis
of performance and correctness of mobile apps. The techniques
presented in this paper are mainly related to correctness, but some of
the discussions relate to performance as well. I consider paramount
important to give developers mechanisms to analyse mobile apps way
before they start thinking on their implementation or can make any
decision about the use of any particular technology. Our approach to
the analysis of the design of mobile apps relies on first writing the
software requirements of the app as User Stories, and then formalising
them directly in Event-B. Writing software requirements in Event-B
demand a high level of formality from developers as one should write
invariants and predicates in logic, and should use (semi-) automatic
theorem provers to validate one's understanding of the system that one
has in mind. This paper abstract away from issues related to the use
of Rodin to discharge Proof Obligations. Writing invariants requires
certain discipline and training, but then, I judge that it's the same
kind of discipline and training a developer would require to
appropriate and master every new technology of interest.

The EventB2Java tool generates Java class implementations for
events. EventB2Java blindly translates event actions into Java, hence,
if a set or relation expression occurs repeated several times in the
right-hand side of an assignment, EventB2Java translates it several
times to Java; the tool does not perform any kind of
preprocessing. Nevertheless, the Java code can serve as a prototype
implementation and be used to animate and check the actual behaviour
of the Event-B model in Java.

In addition to the approach presented in this paper to check the
design of Android apps, one can use ProB \cite{prob} to check for
deadlock conditions. For instance, the analysis performed in Section
\ref{decisions0} for event \ebcode{chatting} can be supplemented with
the use of ProB to check for deadlocks in the interleaving of
\ebcode{chatting} with all the other machine events.

In what follows I give a list of functionality that though wasn't
included in my model of WhatsApp and that it's worthwhile pursuing as
future work as future work. $(ii.)$ Archiving chats. I consists in
backing up a chat, making it inactive, and hiding it from the
user. $(ii.)$ Pinning a chat. This is a more an interface-related
requirement, it consists in moving a chat up in the list of existing
chats so that the user does not need to scroll down to search for that
chat in her chat list. $(iii.)$ Contact lists. Chats can only be
created out of a local contact list. $(iv.)$ Copying a message or
content. The message or content can be pasted thereafter. $(v.)$ Time
stamps. Users can check the time stamps of sent or received
messages. $(v.)$ Group of users. Users can create groups. These groups
can be used as chat groups or be used in phone calls or video
conferences. $(vii.)$ Phone calls and video conferences. Users can
place phone calls or make video conferences to a person or a group of
persons.

\bibliographystyle{plain}
\bibliography{whatsapp-ESE}

\end{document}